\documentclass[reprint,groupedaddress,unsortedaddress, runinaddress, frontmatterverbose,showkeys,showpacs, preprintnumbers,footinbib, nobibnotes, amsmath, amssymb,floatfix ]{revtex4-1}

\usepackage{graphicx}% Include figure files
\usepackage{dcolumn}% Align table columns on decimal point
\usepackage{bm}% bold math
\usepackage{amsmath}
\usepackage[dvipdfm,colorlinks=true,linkcolor=blue,urlcolor=blue,citecolor=blue]{hyperref}
\begin{document}

%\preprint{ }

\title{\LARGE The manipulated left-handedness in a rare-earth-ion-doped optical fiber by the incoherent pumping field }

\author{Shun-Cai Zhao}
\email[Corresponding author: ]{ zhaosc@kmust.edu.cn }
\affiliation{Department of Physics, Faculty of Science, Kunming University of Science and Technology, Kunming, 650500, PR China}

\author{Hong-Wei Guo}
\affiliation{Department of Physics, Faculty of Science, Kunming University of Science and Technology, Kunming, 650500, PR China}

\author{Xiao-Jing Wei}
\affiliation{Department of Physics, Faculty of Science, Kunming University of Science and Technology, Kunming, 650500, PR China}

\date{\today}

\begin{abstract}
The left-handedness was demonstrated in an $E_{r}^{3+}$-dopped
$Z_{r}F_{4}$-$B_{a}F_{2}$-$L_{a}F_{3}$-$AlF_{3}$-$N_{a}$F (ZBLAFN) optical
fiber modeled by a three-level quantum system. Under the electric and magnetic
components of the probe field driving the transitions of  $^{4}I_{15/2}$-$^{4}I_{13/2}$
and $^{4}I_{9/2}-^{4}I_{13/2}$ in the $E_{r}^{3+}$-dopped optical fiber respectively,
an increasing left-handedness was achieved by the incremental incoherent pumping
field. However, the left-handedness damped when the incoherent pumping field drove the
transition heavily. Our scheme may provide a solid candidate other than the coherent atomic vapour for left-handedness, and may
extend the application of the rare-earth-ion-doped optical fiber in metamaterials
via the external incoherent pumping field.
\end{abstract}

%\pacs{ }

\maketitle

\section{INTRODUCTION}

Optical fibers doped with rare-earth ions, such as $E_{r}^{3+}$,
$N_{d}^{3+}$, $T_{m}^{3+}$, $S_{M}^{3+}$, $H_{o}^{3+}$,
$Y_{b}^{3+}$, and $P_{r}^{3+}$, have attracted significant
scientific and industrial interests due to their applications in
optical fiber amplifiers and fiber lasers\cite{1,2,3}. Especially,
the $E_{r}^{3+}$-doped optical fiber plays a much more important
role in optical fiber communication\cite{4,5,6,7,8,9,10,11}.
And the emission transition $^{4}I_{15/2}- ^{4}I_{13/2}$ in $E_{r}^{3+}$-doped
fiber amplifier is the key element of modern telecommunication systems\cite{4,5}.
In wavelength division multiplexing\cite{6}, the $^{4}I_{13/2}-^{4}I_{15/2}$ emission
and the $^{4}I_{15/2}$-$^{4}I_{13/2}$ absorption transitions of $E_{r}^{3+}$ fulfill
flat emission spectra and wavelength divergence.
The transitions of $^{2}H_{11/2}-^{4}I_{15/2}$ and $^{4}F_{9/2}-^{4}I_{15/2}$ under
infrared radiation excitation can convert fluorescence properties via $E_{r}^{3+}$-doped
nanoparticles of $GdPO_{4}$\cite{12}.

On account of the flexible design and adjustable parameters comparing to their
atomic counterparts, the $E_{r}^{3+}$-doped optical fiber systems play a more
important role in practical application for quantum optics. And some nonlinear
quantum optical phenomena, such as gain leveling\cite{13}, optical bistability
and multi-stability\cite{14}, absorption-amplification response\cite{15},
enhanced index of refraction with vanishing absorption\cite{16} were achieved
in $E_{r}^{3+}$-doped optical fibers recently.

In this work, under the electric and magnetic components of the probe field coupling the transitions of
 $^{4}I_{15/2}$-$^{4}I_{13/2}$ and $^{4}I_{9/2}-^{4}I_{13/2}$, respectively, we theoretically investigate the feasibility
of left-handedness via the incoherence pumping
field in the $E_{r}^{3+}$-dopped ZBLAFN optical fiber.
For the simplify of experimental realization, we simulate the $E_{r}^{3+}$-dopped
optical fiber with an ordinary three-level quantum system, and the
switching from increasing to decreasing left-handedness can be implemented simply
by adjusting the pumping rate of the incoherent pumping field.
In our scheme, a new left-handed material may be explored instead of artificial composite
metamaterials\cite{17}, photonic crystal structures\cite{18}, transmission line
simulation\cite{19} and chiral media \cite{20}, which may extend the mediums for left-handedness
and the application domain for $E_{r}^{3+}$-dopped optical fiber in metamaterials

\section{MODEL AND EQUATION}

The $E_{r}^{3+}$-dopped ZBLAFN optical
fiber can be modeled by three-level quantum system shown in Fig. 1.
The levels $|^{4}I_{15/2}\rangle$, $|^{4}I_{13/2}\rangle$ and $|^{4}I_{9/2}\rangle$ in the $E_{r}^{3+}$
will behaved the $|1\rangle$, $|2\rangle$ and $|3\rangle$ states, respectively. In such a modeled 3-level system,
the parities of levels $|2\rangle$ and $|3\rangle$ are set to be identical, which is opposite to
$|1\rangle$. An incoherent pump field pumps atoms in level $|1\rangle$ into upper level $|3\rangle$
with its pumping rate being $2\Gamma$. The possible optical transition $|1\rangle$ $\leftrightarrow$ $|2\rangle$
is mediated by a weak probe laser field with central frequency $\omega_{e}$ and
Rabi frequency $\Omega_{e}$=$\vec{E}_{e}d_{21}/\hbar$. Because of the parity selection rules, the two levels $|1\rangle$
and $|2\rangle$ with electric dipole element $d_{21}$ =$\langle2|$$\hat{\vec{d}}$$|1\rangle$$\neq0$
are coupled by the electric component of the weak probe field, where $\hat{\vec{d}}$ is the electric dipole operator.
The magnetic component of the probe field with frequency $\omega_{b}$ and Larmor frequency of $\Omega_{b}$=$\vec{B}$$ \mu_{32}/2\hbar$
is applied to the magnetic-dipole transition $|2\rangle$$\leftrightarrow$$|3\rangle$, where $\mu_{32}$ is the corresponding magnetic-dipole matrix element.
Interestingly, the transitions $|1\rangle$ $\leftrightarrow$ $|3\rangle$, $|1\rangle$ $\leftrightarrow$ $|2\rangle$ and $|3\rangle$ $\leftrightarrow$ $|2\rangle$
coupled by the incoherent pump field, the electric and magnetic components of the probe laser field respectively
form a closed-loop configuration.

\begin{figure}[htp]
\center
\includegraphics[totalheight=1.8 in]{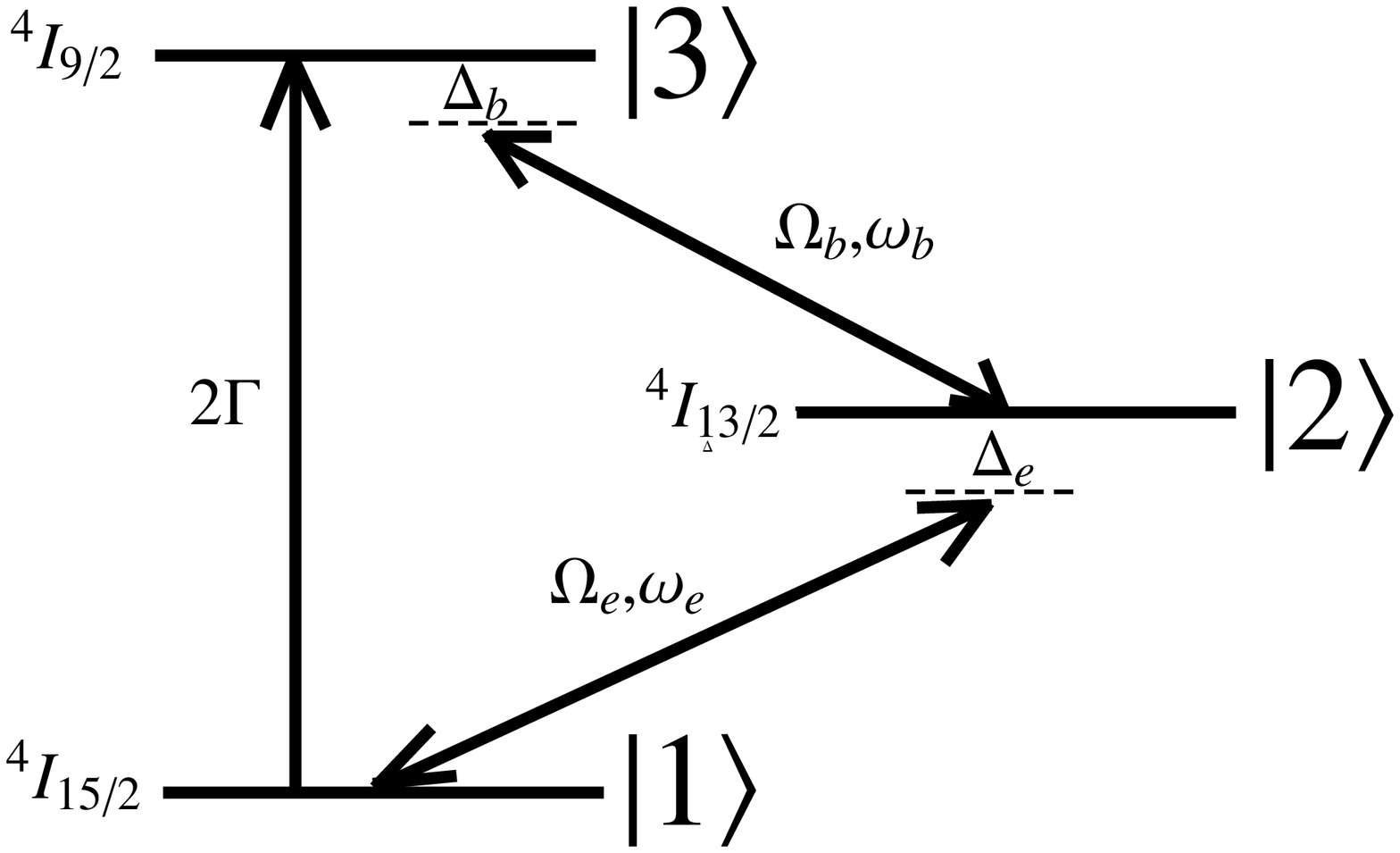 }
\caption{Schematic diagram $E_{r}^{3+}$-dopped $Z_{r}F_{4}$-$B_{a}F_{2}$-$L_{a}F_{3}$-$AlF_{3}$-$N_{a}$F (ZBLAFN) in optical fibers modeled by a three-level system. The levels $^{4}I_{15/2}$, $^{4}I_{13/2}$, and $^{4}I_{9/2}$ behave the $|1\rangle$, $|2\rangle$ and $|3\rangle$ state labels, respectively}
\end{figure}\label{Fig.1}

Then the semi-classical interaction Hamiltonian $H_{int}$ of this ionic system in an $E_{r}^{3+}$-dopped ZBLAFN optical
fiber is given in the interaction picture under the dipole and the rotating wave approximation as follows,

\begin{align}
H_{int}&=\Delta_{e}(|2\rangle\langle2|+|3\rangle\langle3|)+\Delta_{b}|3\rangle\langle3|+(\Omega_{b}|3\rangle\langle2|\nonumber\\
       &+\Omega_{e}|2\rangle\langle1|+H.c.) \label{eq1}
\end{align}
where H.c. means Hermitian conjugation, $\Delta_{e}$= $\omega_{e}$-$\omega_{21}$ and $\Delta_{b}$= $\omega_{b}$-$\omega_{32}$ are the detunings of the electric and magnetic components of the probe laser field to the transitions $\omega_{21}$ and $\omega_{32}$, respectively. Then the equation of
the time-evolution, i.e., the density matrix equations for the system can be described as  \(\frac{d\rho}{dt}=-\frac{i}{\hbar}[H,\rho]+\Lambda\rho \)
in Eq.(2), where $\Lambda\rho$ represents the irreversible decay part of the $E_{r}^{3+}$ ion system.

\begin{align}
i \dot{\rho_{22}}&=-\Omega_{b}\rho_{32}+\Omega_{b}\rho_{23}-\Omega_{e}\rho_{12}+\Omega_{e}\rho_{21}-i \gamma_{21}\rho_{22}\nonumber\\
                 &+i \gamma_{32}\rho_{33},\nonumber\\
i \dot{\rho_{33}}&=-\Omega_{b}\rho_{23}+\Omega_{b}\rho_{32}-i(\gamma_{31}+\gamma_{32})\rho_{33},\label{eq2}\\
i \dot{\rho_{12}}&=-\Omega_{e}(\rho_{22}-\rho_{11})+\Omega_{b}\rho_{13}-(\Delta_{e}+i\frac{\gamma_{21}}{2}+i\Gamma)\rho_{12},\nonumber\\
i \dot{\rho_{13}}&=\Omega_{b}\rho_{12}-\Omega_{e}\rho_{23}-[(\Delta_{e}+\Delta_{b})+i\frac{\gamma_{21}}{2}+i\Gamma]\rho_{13},\nonumber\\
i \dot{\rho_{23}}&=-\Omega_{b}(\rho_{33}-\rho_{22})-\Omega_{e}\rho_{13}-(\Delta_{b}+i\frac{\gamma_{21}+\gamma_{31}+\gamma_{32}}{2})\nonumber\\
                 &\rho_{13},\nonumber
\end{align}

\noindent where $\rho_{ij}$=$\rho^{*}_{ji}$(i,j=1,2,3) and the density matrix elements were
constrained by the conditions: $\rho_{11}+\rho_{22}+\rho_{33}$=$1 $. Here, $\gamma_{ij}$ designates
the decay rates from $|i\rangle$ to $|j\rangle$.

In the classical electromagnetic theory, the electric
polarizability is a rank 2 tensor and defined by its Fourier transform
$\vec{P}_{e}(\omega_{e})$ $=$$\epsilon_{0}$$\alpha_{e}(\omega_{ e})$$\vec{E}(\omega_{e})$, which
can be calculated by the trace computation of the definition $\vec{P}_{e}$
=Tr$\{$${\hat{\rho}\vec{d}}$$\}$=$\rho_{12}d_{21}$+c.c.. Here, in the $E_{r}^{3+}$-dopped
optical fiber we consider the polarizability of the incident field $\vec{E}_{e}$ carrying
out at the frequency $\omega_{e}$. Therefore, we adopt the explicit $\omega_{e}$ dependence
$\alpha_{e}(\omega_{P})\equiv\alpha_{e}$, and $\vec{E}_{e}$ was set to parallel to the atomic dipole $\vec{d}_{21}$ so
$\alpha_{e}$ as to be a scalar. Then its expression can be represented as follows:

\begin{align}
\alpha_{e}=\frac{\vec{d}_{21}\rho_{12}}{\epsilon_{0}\vec{E}_{e}}=\frac{\mid
{d_{21}}\mid^{2} \rho_{12}}{\epsilon_{0}\hbar\Omega_{e}},
\end{align}

The classical magnetic polarizations in the $E_{r}^{3+}$-dopped
optical fiber can be achieved in the same way, i.e.,
$\vec{P}_{b}(\omega_{b})$=$\mu_{0}\alpha_{b}\vec{B}(\omega_{b})$,
which can be obtained  by the mean value of the atomic dipole moment operator
via $\vec{P}_{b}$=Tr$\{$${\hat{\rho}\vec{\mu}}$$\}$=$\rho_{32}$$\mu_{23}+c.c$.
For the simplification, we choose magnetic dipole is perpendicular to the induced
electric dipole in accordance with the classical Maxwell's electromagnetic wave-vector
relation. Then the magnetization  $\alpha_{m}$ is scalar, and its expression is as follows:

\begin{align}
\alpha_{m}=\frac{\mu_{0}\vec{\mu}_{32}\rho_{23}}{\vec{B}}=\frac{\mu_{0}\mid\mu_{32}\mid^{2}\rho_{23}}{\hbar\Omega_{B}}.
\end{align}

The relative permittivity and relative permeability of the $E_{r}^{3+}$-dopped
optical fiber can be given according to the Clausius-Mossotti relations considering
the local effect in dense medium\cite{21,22} as follows:

\begin{align}
\epsilon_{r}=\frac{1+\frac{2}{3}N\alpha_{e}}{1-\frac{1}{3}N\alpha_{e}},
\mu_{r}=\frac{1+\frac{2}{3}N\gamma_{m}}{1-\frac{1}{3}N\gamma_{m}}.\label{eq5}
\end{align}

In the above, the expressions for the electric
permittivity and magnetic permeability of this $E_{r}^{3+}$-dopped
optical fiber were obtained. In the section that follows, we will discuss
its left-handedness via the permittivity, permeability and refractive
index.

\section{RESULTS AND DISCUSSION}

With the steady solutions of Eq.(2) for $\rho_{12}$ and $\rho_{23}$, we can show the numerical results for the electric
permittivity $\varepsilon_{r}$, magnetic permeability $\mu_{r}$ and refractive index n in the $E_{r}^{3+}$-dopped optical fiber.
Before the calculation, some typical parameters should be selected.
In the following numerical calculations, all the parameters will be scaled by $\gamma_{21}$= 90.6 $s^{-1}$ \cite{23}, and we choose the decay rates as $\gamma_{31}$=1.19$\gamma_{21}$ and $ \gamma_{32}$=0.31$\gamma_{21}$ from Ref.\cite{23}.
And we choose the average density for the $E_{r}^{3+}$ ions as N$\approx$$1.04\times10^{20}$$m^{-3}$,
the electric transition dipole moment from $|2\rangle$ $\leftrightarrow$ $|1\rangle$
is chosen as $ d_{21}=2.335\times1.602\times10^{-19}C \cdot m$ \cite{24} and the typical magnetic transition
dipole moment is chosen as $ \mu_{32}=7.0\times10^{-23}Cm^{2}s^{-1}$\cite{25,26}.
The Rabi frequency of the electric component of the probe laser field is set as $\Omega_{e}$=0.5$\gamma_{21}$
with $\Delta_{b}$=0.25$\gamma_{21}$.

\begin{figure}[htp]
\center
\includegraphics[totalheight=1.8 in]{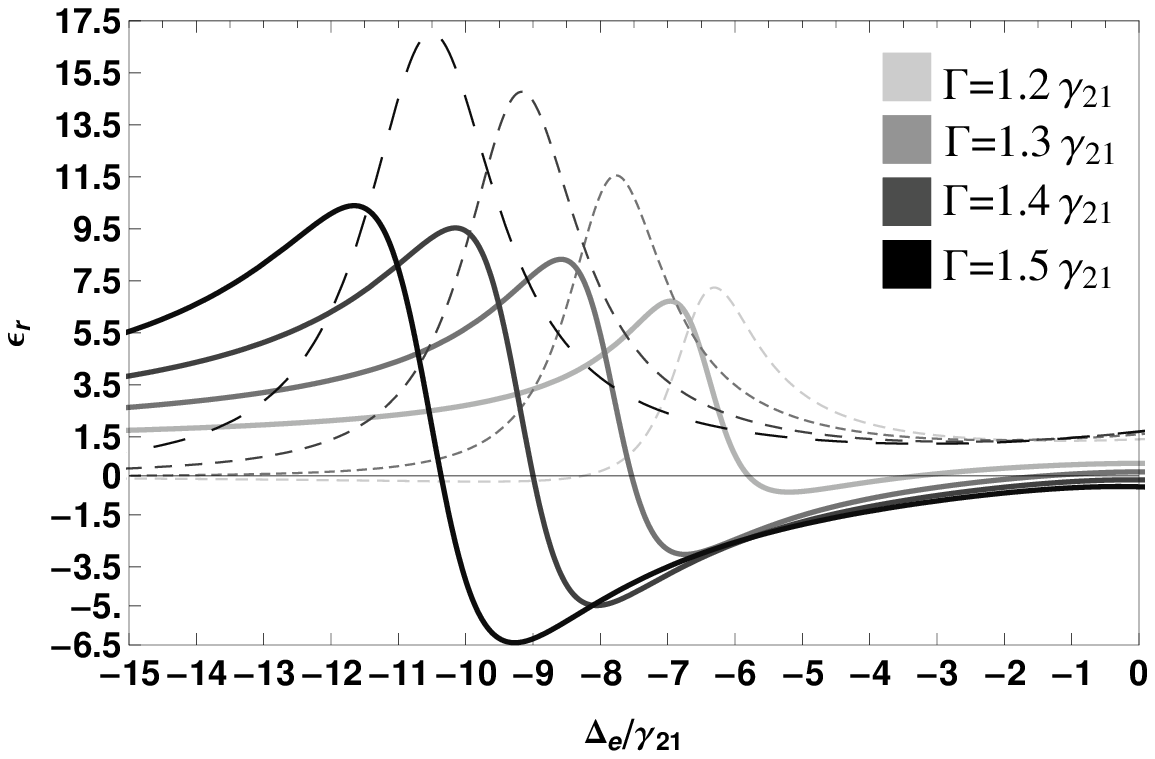 }
\caption{ Real (solid lines) and imaginary (dashed lines) parts of permittivity $\varepsilon_{r}$ as a function of the rescaled detuning parameter $\Delta_{e}/\gamma_{21}$ for $\Gamma$=$1.2\gamma_{21}$, $\Gamma$=$1.3\gamma_{21}$, $\Gamma$=$1.4\gamma_{21}$, $\Gamma$=$1.5\gamma_{21}$.}
\end{figure}\label{Fig.2}

According to the refraction definition of the left-handed material \(n =-\sqrt{\varepsilon_{r}\mu_{r}}\)\cite{27}, we plot the permittivity $\varepsilon_{r}$, permeability  $\mu_{r}$ and refractive index n versus $\Delta_{e}/\gamma_{21}$ with different incoherent pumping frequencies \(\Gamma\) in Fig.2, 3 and 4. The coincide intervals for negative $Re[\varepsilon_{r}]$ and $Re[\mu_{r}]$ will demonstrate the left-handedness in the $E_{r}^{3+}$-dopped optical fiber. In Fig.2, we notice the value and interval for negative $Re[\varepsilon_{r}]$ are increasing when the incoherent pumping frequencies are varied by $\Gamma$=$1.2\gamma_{21}$, $\Gamma$=$1.3\gamma_{21}$, $\Gamma$=$1.4\gamma_{21}$, $\Gamma$=$1.5\gamma_{21}$. Which demonstrates the incoherent pump field coupling level $|1\rangle$ into upper level $|3\rangle$ can incur the creasing negative $Re[\varepsilon_{r}]$. On the same parametric condition \(\mu_{r}\) was plot in Fig.3. It's noted that $Re[\mu_{r}]$ maintains negative value in the interval of [-15\(\gamma_{21}\), 0] when the incoherent pumping field modulates its frequency, which shows the $E_{r}^{3+}$-dopped optical fiber has negative $Re[\mu_{r}]$
in the same intervals as negative $Re[\varepsilon_{r}]$, and demonstrates the left-handedness in the $E_{r}^{3+}$-dopped optical fiber.

\begin{figure}[htp]
\center
\includegraphics[totalheight=1.8 in]{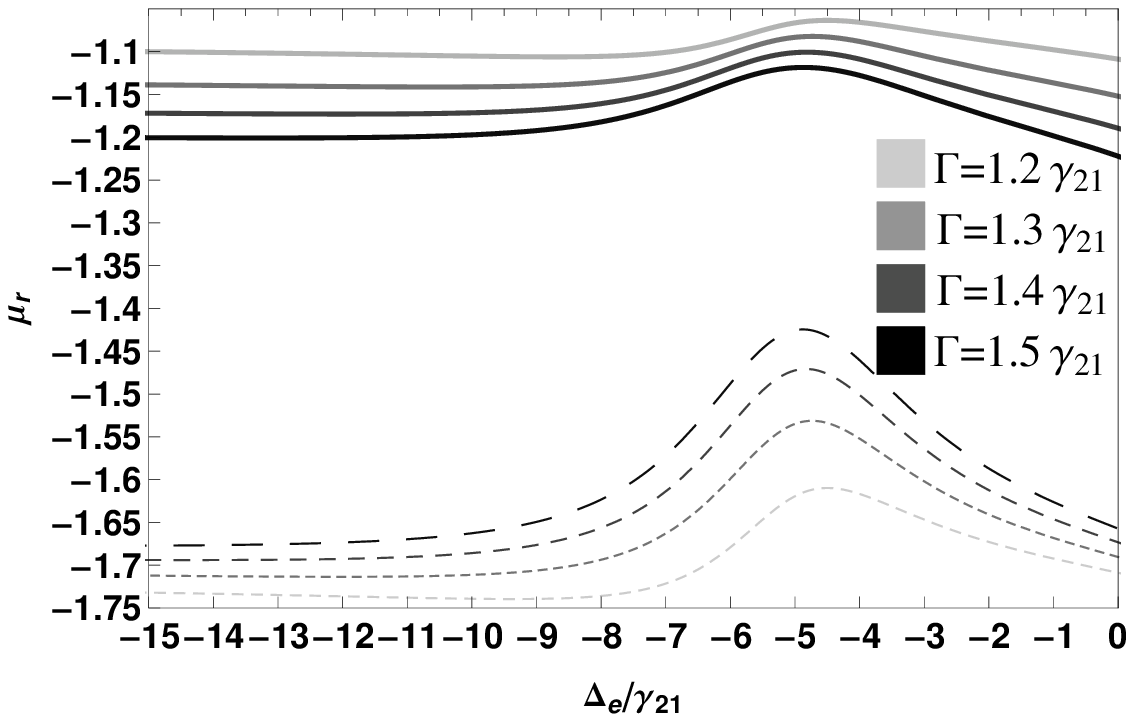 }
\caption{ Real (solid lines) and imaginary (dashed lines) parts of permeability  $\mu_{r}$ as a function of the rescaled detuning parameter $\Delta_{e}/\gamma_{21}$ for $\Gamma$=$1.2\gamma_{21}$, $\Gamma$=$1.3\gamma_{21}$, $\Gamma$=$1.4\gamma_{21}$, $\Gamma$=$1.5\gamma_{21}$, and the other parameters are same as in Fig.2.}
\end{figure}\label{Fig.3}
The plots for $Re[n]$ in Fig.4 also demonstrate this. Fig.4 shows that $Re[n]$ maintain negative value in the interval of [-15\(\gamma_{21}\), 0] and its values are gradually enlarging when the incoherent pumping field is modulated in the same way as Fig.2.

The reason may come from the growing incoherent pumping field, which drives the transition $|3\rangle$ $\leftrightarrow$ $|1\rangle$ and brings out the changing populations between $|3\rangle$ and $|1\rangle$ in the $E_{r}^{3+}$-dopped optical fiber. The variation of population between $|3\rangle$ and $|1\rangle$ results in the field-induced interference effect on the electric and magnetic polarization, which leads to the increasing left-handedness eventually.

\begin{figure}[htp]
\center
\includegraphics[totalheight=1.8 in]{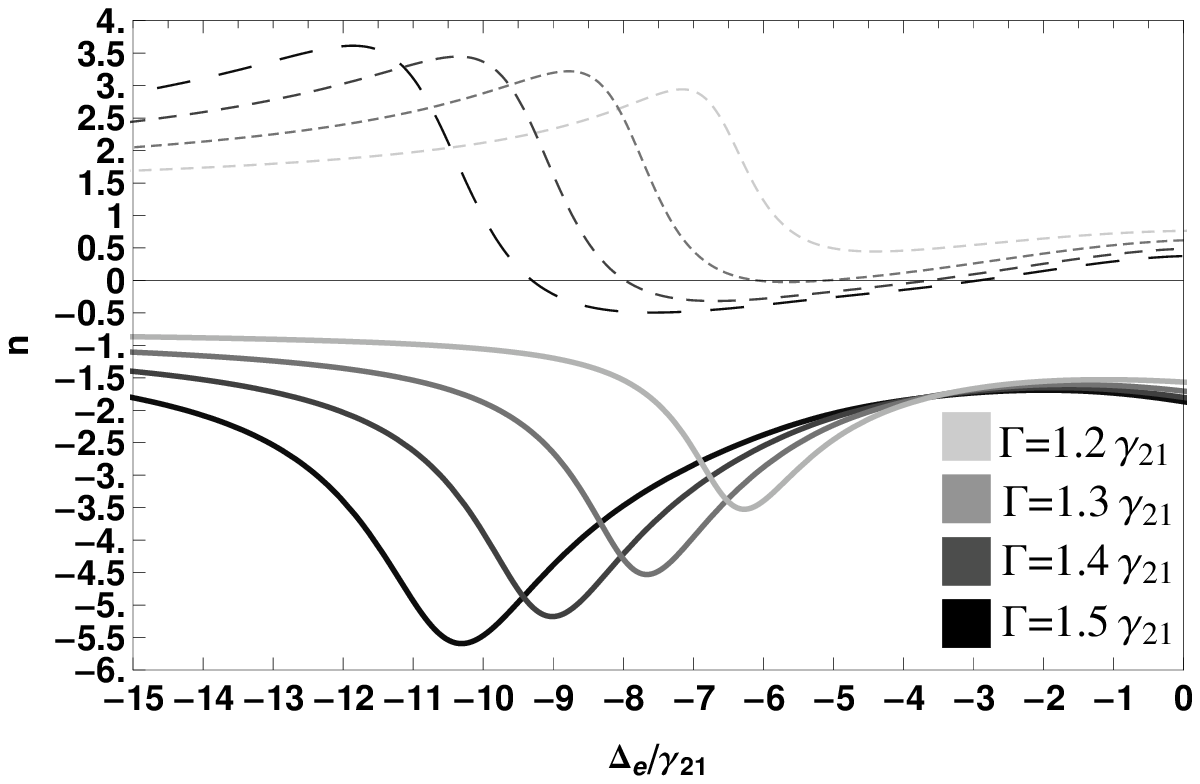 }
\caption{ Real (solid lines) and imaginary (dashed lines) parts of refractive index  $ n $ as a function of the rescaled detuning parameter $\Delta_{e}/\gamma_{21}$ for $\Gamma$=$1.2\gamma_{21}$, $\Gamma$=$1.3\gamma_{21}$, $\Gamma$=$1.4\gamma_{21}$, $\Gamma$=$1.5\gamma_{21}$, and the other parameters are same as in Fig.2.}
\end{figure}\label{Fig.4}

Above all, the incoherent pumping field plays a important role in implementing left-handedness and attracts us mostly. What's the result for the strongly incoherent pumping on the transition $|3\rangle$ $\leftrightarrow$ $|1\rangle$ $?$ In the following, the incoherent pumping rate was set as $\Gamma$=$10\gamma_{21}$, $\Gamma$=$15\gamma_{21}$, $\Gamma$=$20\gamma_{21}$, $\Gamma$=$25\gamma_{21}$ in Fig.5, 6 and 7, which are more stronger than those utilized before. As shown in Fig.5, 6, we note that the intervals for simultaneous negative $Re[\varepsilon_{r}]$ and $Re[\mu_{r}]$ are $[0,-15]$, which means a wide adjustable parameters for implementing left-handedness when the incoherent pumping field drives $|3\rangle$ $\leftrightarrow$ $|1\rangle$ heavily. However, the increasing  pumping rate $\Gamma$ leads to the damping values for $Re[\varepsilon_{r}]$ and $Re[n]$ in Fig.5 and Fig.7, even so the slightly rising $Re[\varepsilon_{r}]$ in Fig.6. As demonstrates the decreasing left-handedness when the incoherent pumping field drives  $|3\rangle$ $\leftrightarrow$ $|1\rangle$ heavily. It means the strong incoherent pumping field plays a destructive role in implementing left-handedness in the $E_{r}^{3+}$-dopped optical fiber.
\begin{figure}[htp]
\center
\includegraphics[totalheight=1.8 in]{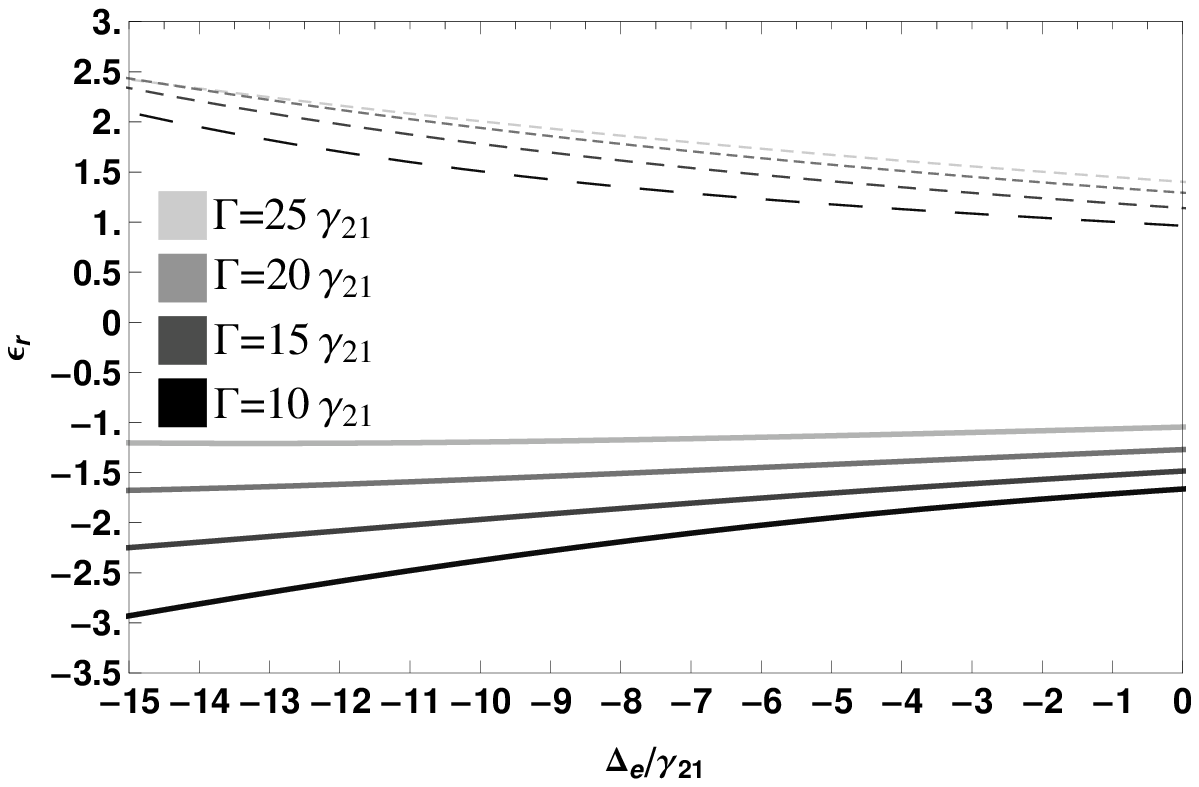 }
\caption{ Real (solid lines) and imaginary (dashed lines) parts of permittivity \(\varepsilon_{r}\) as a function of the rescaled detuning parameter \(\Delta_{e}/\gamma_{21}\) for $\Gamma$=$10\gamma_{21}$, $\Gamma$=$15\gamma_{21}$, $\Gamma$=$20\gamma_{21}$, $\Gamma$=$25\gamma_{21}$, and the other parameters are same as in Fig.2.}
\end{figure}\label{Fig.5}

\begin{figure}[htp]
\center
\includegraphics[totalheight=1.8 in]{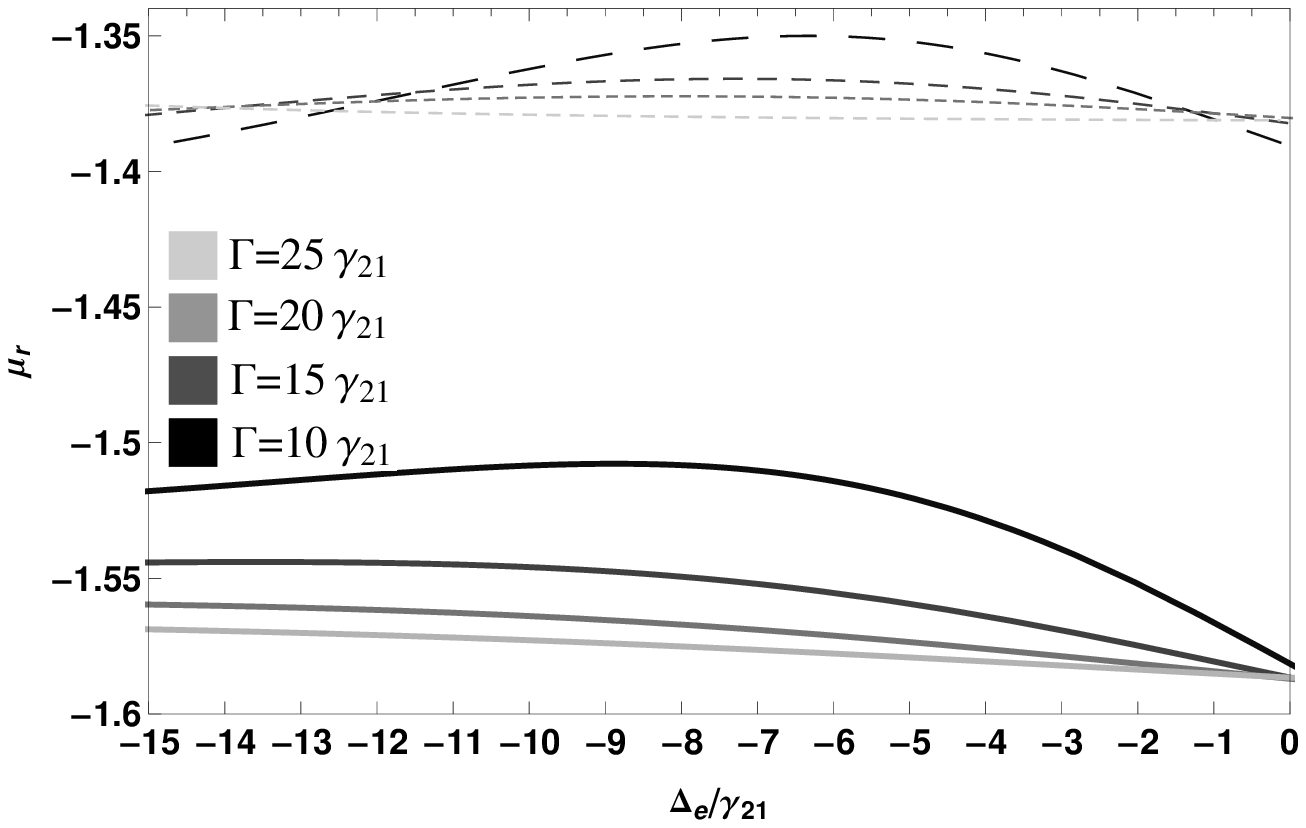 }
\caption{ Real (solid lines) and imaginary (dashed lines) parts of permeability \(\mu_{r}\) as a function of the rescaled detuning parameter $\Delta_{e}/\gamma_{21}$ for $\Gamma$=$10\gamma_{21}$, $\Gamma$=$15\gamma_{21}$, $\Gamma$=$20\gamma_{21}$, $\Gamma$=$25\gamma_{21}$, and the other parameters are same as in Fig.2.}
\end{figure}\label{Fig.6}

From Fig.4 and Fig.7, we conclude that the different intensities of the incoherent pumping field result in the increasing or damping left-handedness in the $E_{r}^{3+}$-dopped optical fiber. The reason may qualitatively explain as the $E_{r}^{3+}$ ion's closed-loop configuration. When the incoherent pumping field drives transition $|1\rangle$ $\leftrightarrow$ $|3\rangle$ heavily, it influences the interferences between $|3\rangle$ $\leftrightarrow$ $|2\rangle$ and $|1\rangle$ $\leftrightarrow$ $|2\rangle$ simultaneously. The diminished quantum coherence and interference decrease the left-handedness and lead to the shrinking of $Re[n]$ in the $E_{r}^{3+}$-dopped optical fiber.

\begin{figure}[htp]
\center
\includegraphics[totalheight=1.8 in]{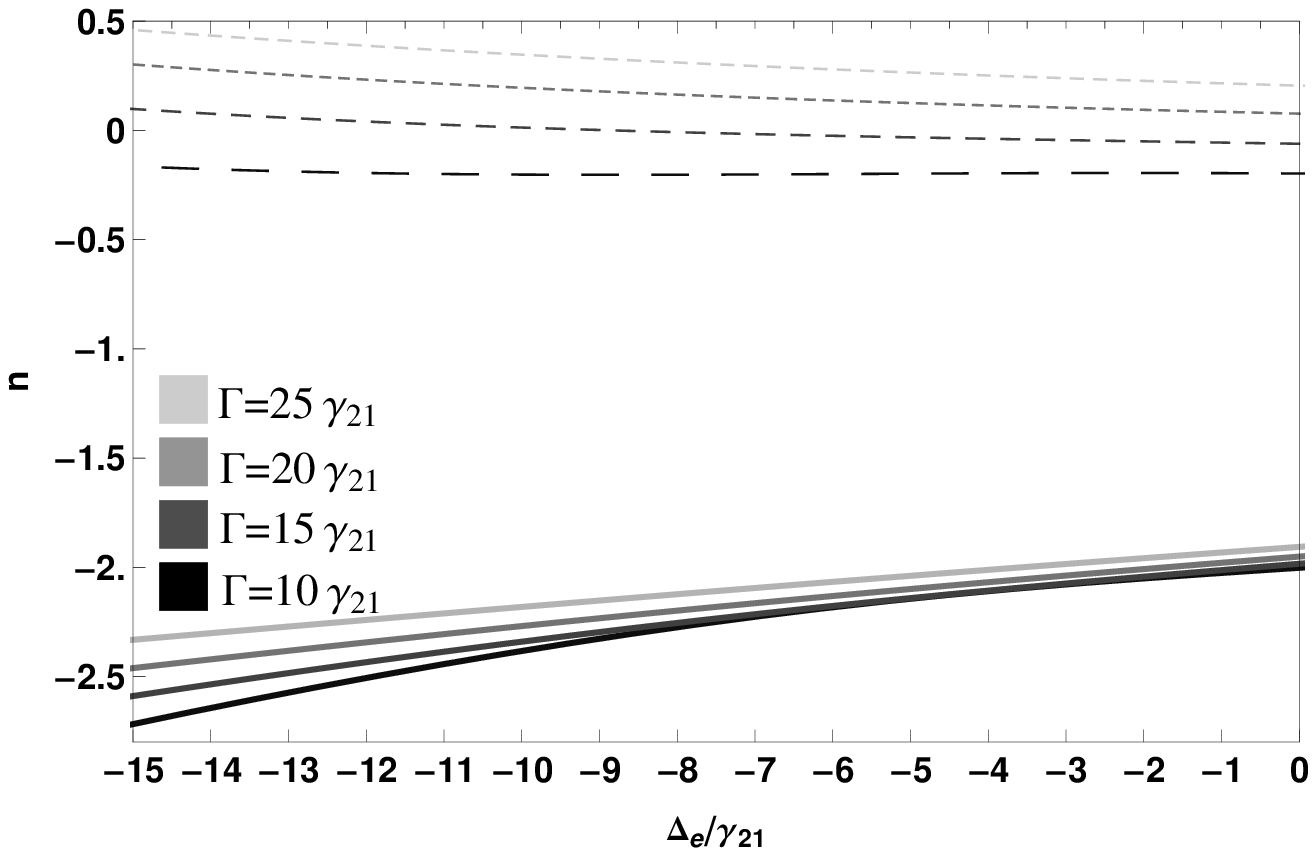 }
\caption{ Real (solid lines) and imaginary (dashed lines) parts of refractive index  $ n $ as a function of the rescaled detuning parameter $\Delta_{e}/\gamma_{21}$ for $\Gamma$=$10\gamma_{21}$, $\Gamma$=$15\gamma_{21}$, $\Gamma$=$20\gamma_{21}$, $\Gamma$=$25\gamma_{21}$, and the other parameters are same as in Fig.2.}
\end{figure}\label{Fig.7}

In our scheme, we used a three-level atomic system to model the emission transitions $^{4}I_{15/2}- ^{4}I_{13/2}$ and $^{4}I_{13/2}-^{4}I_{15/2}$ in an $E_{r}^{3+}$-dopped ZBLAFN optical fiber. In order to achieve an asymptotic simulation result, the decay rates were set from some relevant research results\cite{23}. Undoubtedly, these parameters utilized the simulating process promote the probability in the coming experiment.

From the above analysis, the left-handedness was demonstrated in an $E_{r}^{3+}$-dopped ZBLAFN optical fiber modeled by a three-level quantum system.
In the literature, there was a quantum optical method, i.e., photonic resonant materials\cite{26} to realize left-handed media.
In which a controllable manipulation of the left-handedness by using the external fields (e.g.,just like the incoherent pumping field in our scheme) interacting with coherent atomic vapour. However, in the current work, we achieved the left-handedness in the solid candidate which is the prominant feature other than the coherent atomic vapour. We would like to point out one can fruit the candidate of $E_{r}^{3+}$-dopped ZBLAFN optical fiber for left-handedness in our scheme.
In the relevant experiment, the the solid candidate may be liable to left-handedness. Meanwhile, the candidate for left-handedness of $E_{r}^{3+}$-dopped ZBLAFN optical fiber paved a solid candidate for left-handedness and extended the application of the rare-earth-ion-doped optical fiber in metamaterials via the external incoherent pumping field, which may attracted some interesting in the future.

\section{CONCLUSION}

In conclusion, our scheme achieved left-handedness in an $E_{r}^{3+}$-dopped
 ZBLAFN optical fiber via the incoherent pumping field driving the transition of $^{4}I_{9/2}-^{4}I_{13/2}$.
Through simulating $E_{r}^{3+}$ ion system with a three-level quantum system, a gradually
growing left-handedness was achieved by the faint growth of the incoherent pumping field.
However, the damping left-handedness emerges when the incoherent pumping field drives the transition strongly.
Our scheme may propose a new avenue for implementing left-handedness and may extend the
applied domain for rare-earth-ion-doped optical fiber in metamaterials.

\begin{acknowledgments}
This work is supported by the National Natural Science Foundation of China ( Grant Nos. 61205205 and 6156508508 ),
the General Program of Yunnan Applied Basic Research Project, China
( Grant No. 2016FB009 ) and the Foundation for Personnel training projects of Yunnan Province, China ( Grant No. KKSY201207068 ).
\end{acknowledgments}

%\appendix*
%\section{appendix}

\end{document}